\documentstyle[12pt,a4,psfig]{article}
\def\eg{{\it e.g.}}
\def\chg{\tilde{\omega}}
\def\cosd{\cos\delta}
\def\kplusdecay{K^+ \to \pi^+ \nu \nubar}

\def\ie{{\it i.e.}}
\def\gev{{\rm GeV} }

\def\vsk#1{\noalign{\vskip#1 cm}}
\def\vsp#1{\vspace{#1 cm}}
\def\ov{\overline}
\def\xd{x_d^{}}
\def\ek{\epsilon_K^{}}

\def\sn2w{\sin^2\theta_W}

\def\lsim{{\mathop <\limits_\sim}}

\def\gm5{\gamma_5}

\def\to{\rightarrow} 
\def\longto{\longrightarrow} 
\def\nubar{\ov{\nu}}
\def\bbbar{B^0\mbox{-}\ov{B^0}}
\def\kkbar{K^0\mbox{-}\ov{K^0}}
\def\etal{{\it et al.}}
\newcommand{\beq}{\begin{equation}}
\newcommand{\eeq}{\end{equation}}
\newcommand{\bea}{\begin{eqnarray}}
\newcommand{\eea}{\end{eqnarray}}
\newcommand{\bsub}{\begin{subequations}}
\newcommand{\esub}{\end{subequations}}

\renewcommand{\theequation}{\thesection.\arabic{equation}}
\newcommand{\clean}{\setcounter{equation}{0}}

\def\PRD#1#2#3{Phys. Rev. {\bf D#1} (19#2) #3}
\def\NPB#1#2#3{Nucl. Phys. {\bf B#1} (19#2) #3}
\def\ZPC#1#2#3{Z. Phys. {\bf C#1} (19#2) #3}
\def\PLB#1#2#3{Phys. Lett. {\bf B#1} (19#2) #3}
\def\PRL#1#2#3{Phys. Rev. Lett. {\bf #1} (19#2) #3}
\def\PRep#1#2#3{Phys. Rep. {\bf #1} (19#2) #3}
\def\PTP#1#2#3{Prog. Theor. Phys. {\bf #1} (19#2) #3}
\def\MPL#1#2#3{Mod. Phys. Lett. {\bf A#1} (19#2) #3}

\def\figmssm{
\begin{figure}[t]
\begin{center}
\leavevmode\psfig{figure=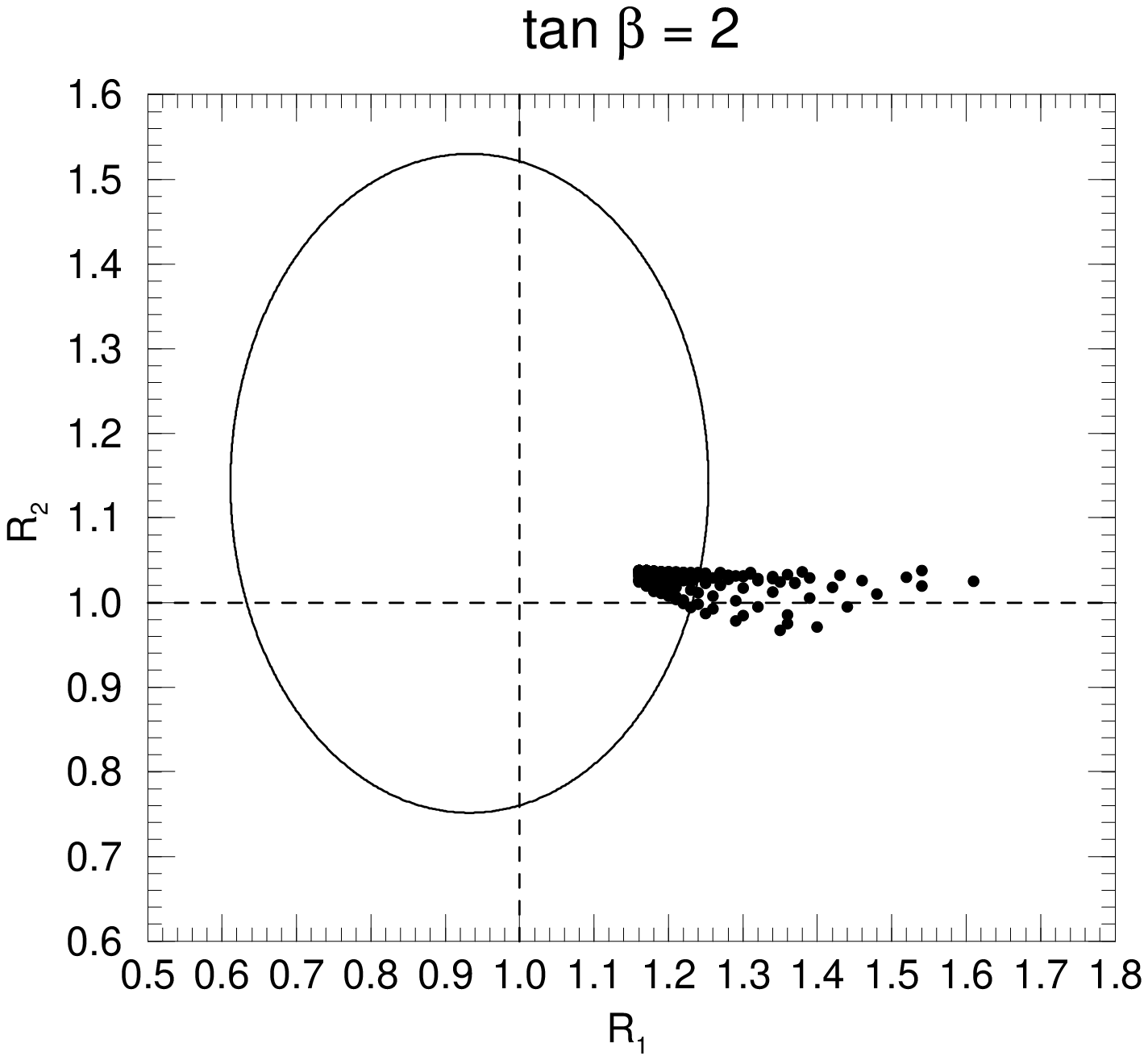,width=6cm}
\leavevmode\psfig{figure=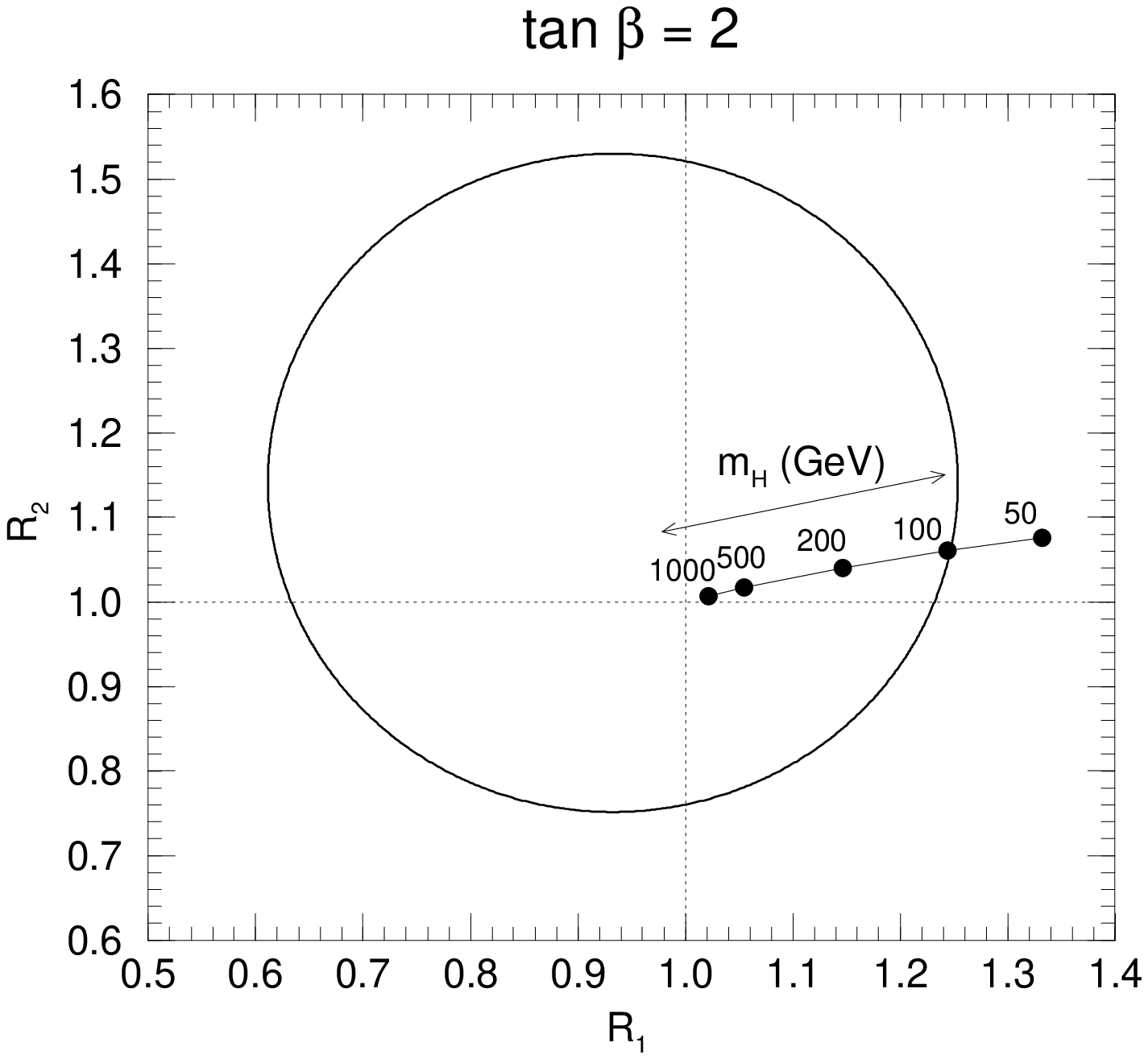,width=6cm}
\end{center}
\caption{
The MSSM (left) and THDM (right) contributions 
to $R_1$ and $R_2$ parameters for 
$\tan\beta = 2$. 
The 1-$\sigma$ allowed region of $R_1$ and $R_2$ 
parameters for $\cosd = 0.36$ is also shown. 
}
\label{r2_mssm}
\vsp{0.5}
\end{figure}
}
\def\figbbkkonesigma{
\begin{figure}[t]
\begin{center}
\leavevmode\psfig{figure=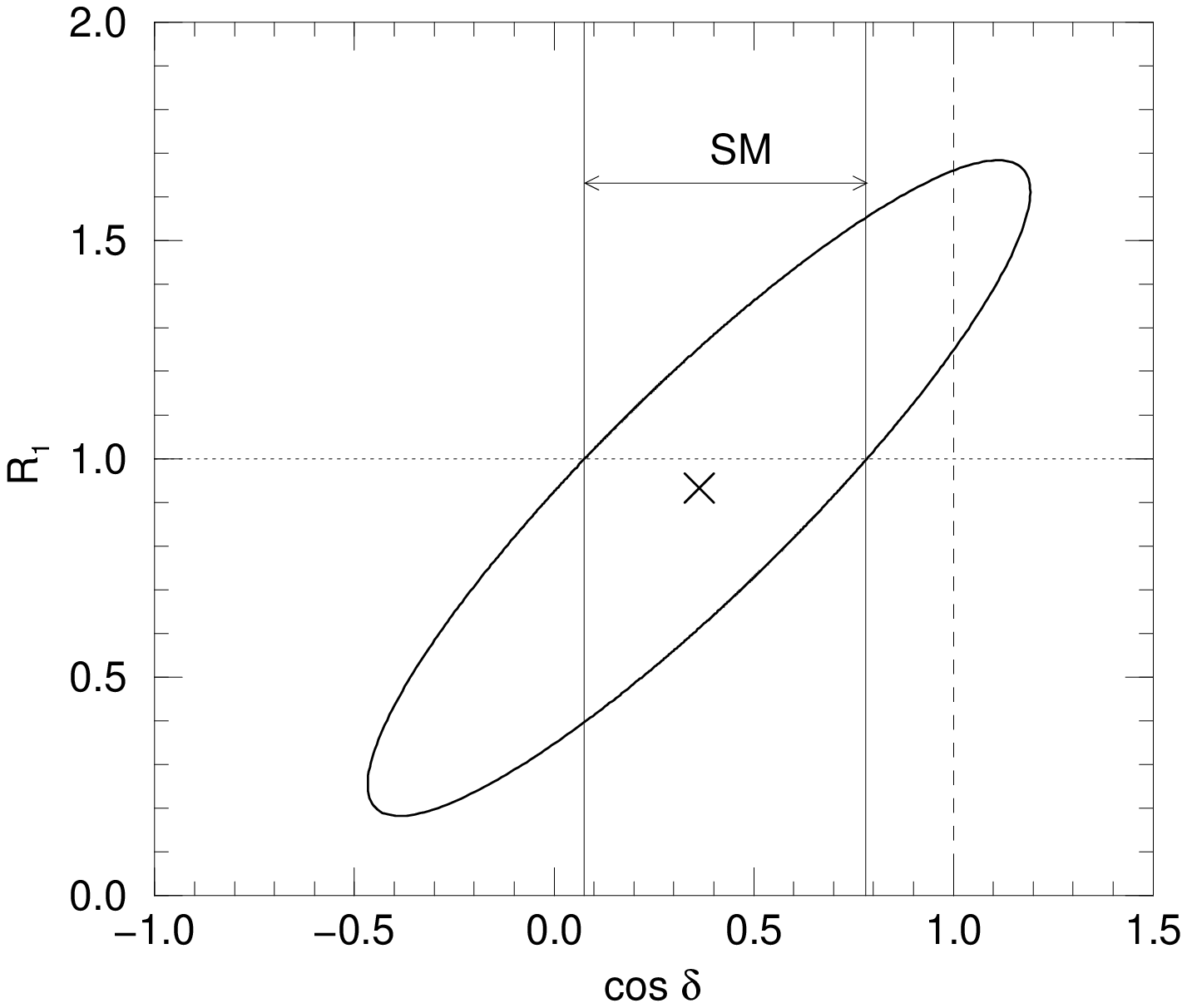,width=9cm}
\end{center}
\caption{ 
The 1-$\sigma$ (39\% CL) allowed region from the experimental 
results of the $\bbbar$, $\kkbar$ mixings. 
The range between the two solid lines is the allowed region 
of $\cosd$ in the SM. 
}
\label{bbkk39cl}
\vsp{0.5}
\end{figure}
}
\def\figsumr1r2{
\begin{figure}[t]
\begin{center}
\leavevmode\psfig{figure=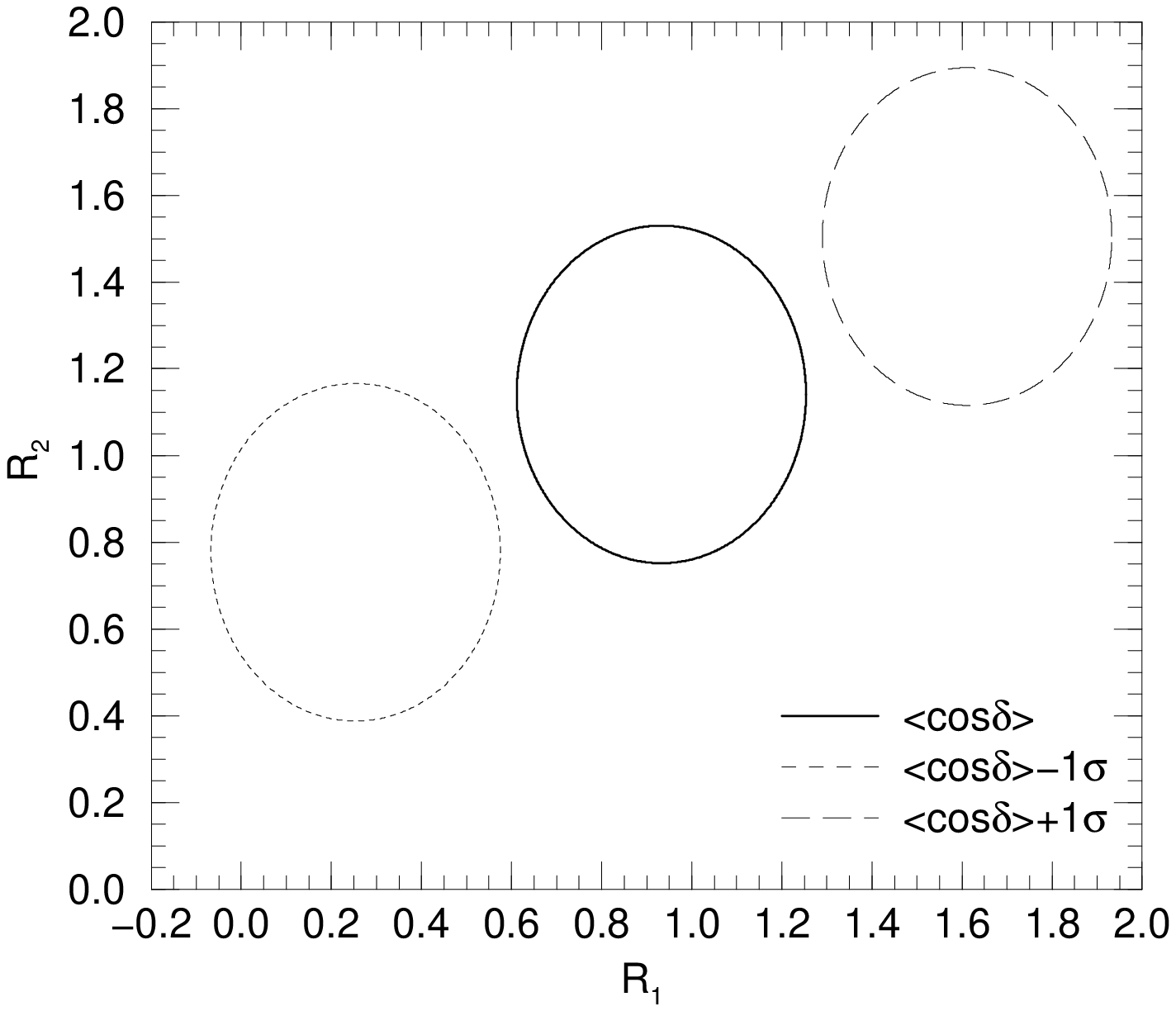,width=9cm}
\end{center}
\caption{The 1-$\sigma$ allowed regions of $R_1,R_2$ 
parameters. Three contours are corresponding to 
$\cosd = 0.36$ (solid line), 
$\cosd = -0.47$ (dotted line) and 
$\cosd = 1.19$ (dashed line), respectively. }
\label{sum_r1r2}
\vsp{0.5}
\end{figure}
}
\makeatletter
\newtoks\@stequation

\def\subequations{\refstepcounter{equation}%
  \edef\@savedequation{\the\c@equation}%
  \@stequation=\expandafter{\theequation}
  \edef\@savedtheequation{\the\@stequation}
  \edef\oldtheequation{\theequation}%
  \setcounter{equation}{0}%
  \def\theequation{\oldtheequation\alph{equation}}}

\def\endsubequations{%
  \ifnum\c@equation < 2 \@warning{Only \the\c@equation\space subequation
    used in equation \@savedequation}\fi
  \setcounter{equation}{\@savedequation}%
  \@stequation=\expandafter{\@savedtheequation}%
  \edef\theequation{\the\@stequation}%
  \global\@ignoretrue}

\def\eqnarray{\stepcounter{equation}\let\@currentlabel\theequation
\global\@eqnswtrue\m@th
\global\@eqcnt\z@\tabskip\@centering\let\\\@eqncr
$$\halign to\displaywidth\bgroup\@eqnsel\hskip\@centering
     $\displaystyle\tabskip\z@{##}$&\global\@eqcnt\@ne
      \hfil$\;{##}\;$\hfil
     &\global\@eqcnt\tw@ $\displaystyle\tabskip\z@{##}$\hfil
   \tabskip\@centering&\llap{##}\tabskip\z@\cr}

\makeatother

\begin{document}
\vspace*{-15mm}
\baselineskip 10pt
\begin{flushright}
\begin{tabular}{l}
{\bf KEK-TH-568}\\
{\bf hep-ph/9804327}\\
{\bf April 1998}
\end{tabular}
\end{flushright}
\baselineskip 18pt 
\vglue 15mm 
\begin{center}
\baselineskip 24pt
\def\thefootnote{\fnsymbol{footnote}}
\setcounter{footnote}{1}
{\Large\bf
Impacts on searching for signatures of new physics 
from $\kplusdecay$ decay}\footnote{Talk given at 
	the workshop on ``Fermion Mass and CP Violation'', 
	Hiroshima, Japan, 5-6 March 1998. }
\vspace{5mm}

{\bf
Gi-Chol Cho\footnote{Research Fellow of the Japan Society 
for the Promotion of Science}
}

\vspace{5mm}
\def\thefootnote{\arabic{footnote}}
\setcounter{footnote}{0}
{\it Theory Group, KEK, Tsukuba, Ibaraki 305, Japan}
\vspace{20mm}
\end{center}
\begin{center}
{\bf Abstract}\\[10mm]
\begin{minipage}{12cm}
\noindent
\baselineskip 14pt
Impacts on new physics search from $\bbbar, \kkbar$ 
mixings and the rare decay $\kplusdecay$ are discussed. 
We show that, in a certain class of new physics models, 
the extra contributions to those processes can be 
parametrized by its ratio to the standard model (SM) 
contribution with the common CKM factors. 
We introduce two ratios to measure the new physics 
contributions, $R_1$ for $\xd$ and $\ek$ parameters, 
and $R_2$ for $\kplusdecay$ decay. 
Then, the experimentally allowed region for the new physics 
contributions can be given in terms of $R_1, R_2$ and the CP 
violating phase $\delta$ of the CKM matrix. 
We find constraints on $R_1$ and $\cosd$ by taking 
account of current experimental data and theoretical 
uncertainties on $\bbbar$ and $\kkbar$ mixings. 
We also study impacts of future improved measurements 
on ($R_1,R_2,\cosd$) basis. 
As typical examples of new physics models, 
we examine contributions to those processes in the minimal 
supersymmetric SM and the two Higgs doublet model. 
\end{minipage}
\end{center}
\newpage
\baselineskip 14pt 
\section{ Introduction }
Processes mediated by flavor changing neutral current (FCNC) 
have been considered as good probes of physics beyond the 
standard model (SM). 
By using the experimentally well measured processes, 
it is expected to obtain an indirect evidence or 
constraints on new physics models.  
An existence of new physics may arise as violation of 
the unitarity of the Cabbibo-Kobayashi-Maskawa (CKM) matrix.  
Such signatures of new physics will be explored through 
the determination of the unitarity triangle 
at B-factories at KEK and SLAC in the near future. 

Typical FCNC processes which have been often used to study 
the new physics contributions are $\bbbar$ and $\kkbar$ mixings. 
Parameters $\xd$ in $\bbbar$ mixing and $\ek$ in $\kkbar$ mixing 
are dominated by the short distance physics and have been 
calculated in the SM and many new physics models. 
Experimentally, both parameters have been measured as~\cite{PDG}
\bsub
\bea
\xd   &=& 0.73 \pm 0.05, \\
|\ek| &=& (2.23\pm 0.013) \times 10^{-3}. 
\eea
\label{eq:exp_values_bbkk}
\esub
\\
On the other hand, there are large theoretical uncertainties 
on both parameters which come from the evaluation of the 
hadronic matrix elements of those processes. 
They are parametrized in terms of decay constants and 
bag-parameters of $B$ or $K$ mesons. 
Thus, the loss of information on the CKM matrix 
elements or the new physics contributions from those processes 
is not avoidable in the level of these uncertainties.

The rare decay $K^+ \to \pi^+ \nu \nubar$ is one of the most 
promising processes to extract clean informations about 
the CKM matrix elements~\cite{buchalla-buras} because 
the decay rate of this process has small theoretical uncertainties. 
The reason can be summarized as: 
\begin{itemize}
\item 	The process is dominated by the short-distance 
	physics. The long-distance contributions have been 
	estimated as $10^{-3}$ smaller than the short-distance 
	contributions~\cite{long_distance}. 
\item   The hadronic matrix element of the decay rate can be 
	evaluated by using that of $K^+ \to \pi^0 e^+ \nu_e$ 
	process which is accurately measured.  
\item   The short distance contributions in the SM have been 
	calculated in the next-to-leading order (NLO) level 
	(for a review, see ~\cite{buchalla-buras}). 
\end{itemize}
With these attractive points, the first observation 
of an event consistent with this decay process which was 
reported by E787 collaboration~\cite{E787} 
\beq
{\rm Br} (K^+ \to \pi^+ \nu \nubar) 
= 4.2^{+9.7}_{-3.5} \times 10^{-10}, 
\eeq
motivates us to examine the implication of 
the above estimate of the branching fraction and of 
its improvement in the near future. 

In this report, we discuss impacts on the search for a new 
physics signal from the above FCNC processes -- $\bbbar, \kkbar$ 
mixings and $\kplusdecay$ decay. 
We focus on a class of new physics models which 
satisfy the following three conditions:
\renewcommand{\labelenumi}{\theenumi}
\renewcommand{\theenumi}{(\roman{enumi})}
\begin{enumerate}
\item	FCNC in the new physics sector is described by 
	the $V-A$ type operator. 
\item 	The flavor mixing in the new physics sector is governed 
	by the SM CKM matrix elements. 
\item 	The net contributions are proportional to the CKM 
	matrix elements which are concerned with the third 
	generation. 
\end{enumerate}

In the following, we will show that new physics contributions 
to those processes can be parametrized by two quantities, $R_1$ for 
$\bbbar,\kkbar$ mixings and $R_2$ for $\kplusdecay$ decay. 
Both quantities are defined as the ratio of the new physics 
contribution to that of the SM. 
Constraints on the new physics 
contributions are summarized in terms of $R_1$, $R_2$ and $\cosd$, 
where $\delta$ is the CP violating phase of the CKM matrix in the 
standard parametrization~\cite{PDG}. 
Taking account of current experimental data on 
$\xd$ and $\ek$ parameters in $\bbbar$ and $\kkbar$ mixings,  
and uncertainties in the hadronic parameters, 
we will show constraints on $R_1$ and $\cosd$. 
In order to see that how $\kplusdecay$ decay could give 
impacts on new physics search, 
we will find constraints on $R_1$, $R_2$ and $\cosd$ 
by assuming the future improvement of the 
Br($\kplusdecay$) measurements. 
As examples of new physics models which naturally satisfy 
the above three conditions, 
we will find the consequences of the minimal supersymmetric 
standard model (MSSM)~\cite{SUGRA} and the two Higgs doublet 
model (THDM)~\cite{higgs_hunter}.

\section{New physics contributions to the FCNC processes 
	in the $B$ and $K$ meson systems}
\clean
The effective Lagrangian for the $\kplusdecay$ process in the SM 
is given by~\cite{inami-lim}: 
\beq
{\cal L}^{K^+}_{eff} = \frac{G_F}{\sqrt{2}} \frac{2\alpha(m_Z)}{\pi} 
	\frac{1}{\sin^2 \theta_W} 
	\ov{\nu}_\ell \gamma^\mu P_L \nu_\ell\ 
	\ov{s}\gamma_\mu P_L d\ 
	\sum_{i = 2,3} V_{i2}^* V_{i1}^{}\ 
	\eta_i D_W^{}(i),
\label{lagrangian_kdecay}
\eeq
where $i$ and $\ell$ are the generation indices for the 
up-type quarks and leptons, respectively. 
The CKM matrix element is given by $V_{ij}$ and 
the projection operator $P_L$ is defined as 
$P_L \equiv (1-\gamma_5^{})/2$. 
The loop function for the $i$-th generation quark 
is denoted by $D_W(i)$ and its explicit form can be 
found in \cite{inami-lim}. 
The QCD correction factor for the top-quark exchange 
has been estimated as $\eta_3 = 0.985$ for $170~{\rm GeV} 
\le m_t^{} \le 190~{\rm GeV}$~\cite{buchalla}.
The QCD correction factor for the charm-quark exchange with 
its loop function is numerically given as $ \eta_2^{} D_W(2) = 
\lambda^4\times(0.40 \pm 0.06 )$~\cite{buchalla2} 
where $\lambda \equiv |V_{12}|$.
The error is due to uncertainties in 
the charm quark mass and higher order QCD corrections. 
Then, summing up the three generations of neutrino, 
the branching ratio is expressed as~\cite{marciano}
\bea
{\rm Br}(K^+ \rightarrow \pi^+ \nu \ov{\nu}) 
&=& 1.57 \times 10^{-4}  
\biggl | V_{32}^* V_{31}^{} \eta_3^{} D_W^{}(3) 
	+ V_{22}^* V_{21}^{}  \eta_2^{} D_W^{}(2)\biggr|^2. 
\label{br_kdecay}
\eea
With the above estimates for the loop functions and the QCD 
correction factors, the branching ratio 
is predicted to be~\cite{recent_estimation}
\beq
{\rm Br}(K^+ \rightarrow \pi^+ \nu \ov{\nu})_{\rm SM} = 
(9.1 \pm 3.8)\times 10^{-11}
\eeq
in the SM, where the error is dominated by the 
uncertainties of the CKM matrix elements. 

The effective Lagrangian of the $\bbbar$ mixing in the SM 
is expressed by 
\bea
{\cal L}^{\Delta B =2}_{eff} 
	= \frac{G_F^2 M_W^2}{4\pi^2}~ 
	\ov{d}\gamma^\mu P_L b~ \ov{d}\gamma_\mu P_L b 
	\sum_{i,j=2,3} 
	V_{i1}^* V_{i3}^{} V_{j1}^* V_{j3}^{}~ 
	F_V^W(i,j). 
\label{lagrangian_bbmixing}
\eea
Likewise, ${\cal L}^{\Delta S =2}_{eff}$  
for the $\kkbar$ mixing is obtained by replacing $V_{i3}$ 
with $V_{i2}$, and the $b$-quark operators  
with the $s$-quark ones, respectively. 
The explicit form of the loop function $F_V^W(i,j)$ 
is given in \cite{inami-lim}. 
The $B$-meson mixing parameter $\xd$ is defined by 
$\xd \equiv \Delta M_B/ \Gamma_B$, where $\Delta M_B$ and 
$\Gamma_B$ correspond to the $B$-meson mass difference and the 
average width of the mass eigenstates, respectively. 
The mass difference is induced by the above $\Delta B=2$ 
operator~(\ref{lagrangian_bbmixing}) and we can express 
the mixing parameter $\xd$ in the SM as 
\beq
\xd = \frac{G_F^2}{6\pi^2}M_W^2 \frac{M_B}{\Gamma_B} f_B^2 B_B 
	\bigl| V_{31}^* V_{33}^{} \bigr|^2 \eta_B 
	\bigl| F_V^W (3,3) \bigr|, 
\label{xd_bb}
\eeq
where $f_B, B_B$ and $\eta_B$ denote the decay constant of 
$B^0$-meson, the bag parameter of $\bbbar$ 
mixing and the short-distance QCD correction factor, respectively.

The CP-violating parameter $\ek$ in the $\kkbar$ system 
is given by the imaginary part of the same box diagram 
of the $\bbbar$ transition besides the external quark lines. 
We can express the $\ek$ parameter in the SM as 
\bea
\ek &=& -e^{i\pi/4} \frac{G_F^2}{12\sqrt{2}\pi^2} M_W^2 
	\frac{M_K}{\Delta M_K} f_K^2 B_K {\rm Im}\biggl\{ 
	(V_{31}^* V_{32}^{})^2 \eta_{K_{33}} F_V^W(3,3) 
	\nonumber \\
	&& ~~~~~
	+ 
	(V_{21}^* V_{22}^{})^2 \eta_{K_{22}} F_V^W(2,2) 
	+ 
	2(V_{31}^* V_{32}^{} V_{21}^* V_{22}^{}) \eta_{K_{32}} 
	F_V^W(3,2)
	\biggr\}, 
\label{epsilon_k}
\eea 
where $f_K$, $B_K$ and $\eta_{K_{ij}}$ represent the 
decay constant, the bag parameter and the QCD correction 
factors, respectively. 

In theoretical estimation of these quantities, 
non-negligible uncertainties come from the evaluations of the 
QCD correction factors and the hadronic matrix elements.  
In our analysis, we adopt the following values:
\bea
\eta_B = 0.55 \pm 0.01~\cite{bb_qcd_buras},
~~~
\sqrt{B_B}f_B = (220 \pm 40)~{\rm MeV}~\cite{fb_const},  
\label{eq:uncertainty_bb}
\eea
for the $\xd$ parameter, and 
\bea
\left.
\begin{array}{lcl}
\eta_{K_{33}} &=& 0.57 \pm 0.01 \\
\eta_{K_{22}} &=& 1.38 \pm 0.20 \\
\eta_{K_{32}} &=& 0.47 \pm 0.04 
\end{array}
\right \}~\cite{bb_qcd_buras,qcd_kk}, 
~~~
B_K = 0.75 \pm 0.15~\cite{recent_estimation}. 
\label{eq:uncertainty_kk}
\eea
for the $\ek$ parameter. 

Next, we consider the new physics contributions to these 
quantities, Br($\kplusdecay$) (\ref{br_kdecay}), 
$\xd$ (\ref{xd_bb}), and $\ek$ (\ref{epsilon_k}). 
In a class of new physics models which satisfy our three 
conditions, the effective Lagrangians can be obtained by 
replacing $D_W^{}(i)$ with $D^{\rm new}(i)$ in 
(\ref{lagrangian_kdecay}),  and $F_V^W(i,j)$ with $F_V^{\rm new}
(i,j)$ in (\ref{xd_bb}) and (\ref{epsilon_k}). 
Then, the effective Lagrangians of these processes in the 
new physics sector should have the following forms;
\bsub
\bea
{\cal L}^{K^+}_{\rm new} &=& \frac{G_F}{\sqrt{2}} 
	\frac{2\alpha(m_Z^{})}{\pi} 
	\frac{1}{\sin^2 \theta_W} 
	\ov{\nu}\gamma^\mu P_L \nu\ 
	V_{32}^* V_{31}^{}\ 
	\ov{s}\gamma_\mu P_L d\ A^{\rm new}, \\
\vsk{0.2}
{\cal L}^{\Delta B =2}_{\rm new} 
	&=& \frac{G_F^2 M_W^2}{4\pi^2}~ 
	\ov{d}\gamma^\mu P_L b~ \ov{d}\gamma_\mu P_L b~
	(V_{31}^* V_{33}^{})^2~ 
	B^{\rm new}, 
\label{eq:del_b=2}  \\
\vsk{0.2}
{\cal L}^{\Delta S =2}_{\rm new} 
	&=& \frac{G_F^2 M_W^2}{4\pi^2}~ 
	\ov{d}\gamma^\mu P_L s~ \ov{d}\gamma_\mu P_L s~
	(V_{31}^* V_{32}^{})^2~ 
	B^{\rm new}. 
\label{eq:del_s=2}
\eea
\esub
It should be noticed that the new physics contributions to 
the $\Delta B=2$ (\ref{eq:del_b=2}) and 
the $\Delta S=2$ (\ref{eq:del_s=2}) 
processes are expressed by the same quantity $B^{\rm new}$. 

There are two cases in which the effective Lagrangians can be 
given by the above forms. 
First, if the contributions from the first two generations 
do not differ much, \ie, 
\bsub
\bea
D^{\rm new}(2) &\approx& D^{\rm new}(1), \\
F_V^{\rm new}(i,1) &\approx&  F_V^{\rm new}(i,2), 
\eea
\esub
the net contributions from the new physics are written by 
using the unitarity of the CKM matrix as; 
\bsub
\bea
\sum_i V_{i2}^* V_{i1}^{} D^{\rm new}(i) &\approx& 
V_{32}^* V_{31}^{} \bigl \{ D^{\rm new}(3) - 
D^{\rm new}(1) \bigr \}, \\
\vsk{0.2}
\sum_{i,j} V_{i1}^* V_{ik}^{} V_{j1}^* V_{jk}^{} F_V^{\rm new}(i,j) 
&\approx& 
(V_{31}^* V_{3k}^{})^2 \bigl\{ F_V^{\rm new}(3,3) 
+ F_V^{\rm new}(1,1) 
\nonumber \\
\vsk{0.1}
&&~~~~~~~~~~~~
- F_V^{\rm new}(3,1) - F_V^{\rm new}(1,3)
\bigr\}, 
\eea
\label{eq:unitarity_cancellation}
\esub
for $k=2,3$. 
We can now define the parameters $A^{\rm new}$ and 
$B^{\rm new}$ as
\bsub
\bea
A^{\rm new} &\equiv& D^{\rm new}(3) - D^{\rm new}(1), \\
B^{\rm new} &\equiv& F_V^{\rm new}(3,3) + F_V^{\rm new}(1,1) 
	- F_V^{\rm new}(3,1) - F_V^{\rm new}(1,3).
\eea
\esub
Second, if the contributions from both the first two 
generations are negligible as compared with those of the 
3rd generation, 
\ie, 
\bsub
\bea
D^{\rm new}(3) &\gg& D^{\rm new}(1), D^{\rm new}(2), \\
F_V^{\rm new}(3,3) &\gg& 
F_V^{\rm new}(1,j), F_V^{\rm new}(2,j), 
F_V^{\rm new}(3,1), F_V^{\rm new}(3,2), 
\eea
\label{eq:yukawa_cancellation}
\esub
the parameters $A^{\rm new}$ and $B^{\rm new}$ become 
\bsub
\bea
A^{\rm new} &=& D^{\rm new}(3), \\
B^{\rm new} &=& F_V^{\rm new}(3,3). 
\eea
\esub

Now, the effects of the new physics contributions to these 
processes can be evaluated by the following ratios~\cite{
cho_kdecay}\footnote{ 
Similar parametrization was used in \cite{susy_buras}.
In the article, the both ratios were defined as complex 
parameters. 
In that case, there are two additional parameters -- two complex 
phases of these ratios. }
\bsub
\bea
R_1 &=& \frac{F_V^W(3,3) + B^{\rm new}}{F_V^W(3,3)}, 
\label{r1}
\\[3mm]
R_2 &=& \frac{D_W^{}(3) + A^{\rm new}}{D_W^{}(3)}. 
\eea
\esub
Once a model of new physics is specified, we can quantitatively 
estimate its effect in terms of $R_1$ and $R_2$.  
Both parameters converge to unity as the new 
physics contributions are negligible, 
\beq
R_1, R_2 \longto 1~~~~{\rm for}~~~
A^{\rm new}, B^{\rm new} \longto 0.
\eeq
Because constraint on $R_2$ is obtained from Br($\kplusdecay$), 
it can be a negative quantity if the extra contributions 
destructively interfere with that of the SM. 
In the following, we consider the cases where the net 
contributions from the new physics sector do not exceed those 
of the SM: 
$A^{\rm new} <  |D_W^{}(3)|$ and $B^{\rm new} < |F_V^W(3,3)|$.  
Then, we study constraints on $R_1$ and $R_2$ from experimental 
data in the range of $0 < R_1, R_2 < 2$.  
\section{Constraints on new physics contributions to FCNC processes}
\clean
Sizable new physics effects to $\xd,\ek$ and Br($\kplusdecay$) 
can be detected as deviations of $R_1$ and $R_2$ from unity. 
In practice, experimentally measurable quantities are 
products of $R_1$ or $R_2$ by the CKM matrix elements. 
In the standard parametrization of the CKM matrix, 
the uncertainty in the CP-violating phase $\delta$ 
dominates that of the CKM matrix elements~\cite{PDG}. 
Hence, together with $R_1$ and $R_2$, we allow $\cosd$ to be 
fitted by the measurements of $\xd,\ek$ and Br($\kplusdecay$). 
By this reason, constraints on $R_1$ and $R_2$ are correlated 
through $\cosd$. 

We perform the $\chi^2$-fit for two parameters $R_1$ and 
$\cosd$ by using experimental data of $\xd$ and 
$\ek$. 
In the fit, we take into account of the theoretical uncertainties 
which are given in (\ref{eq:uncertainty_bb}), 
(\ref{eq:uncertainty_kk}) and 
\bea
\left.
\begin{array}{ccl}
|V_{12}| &=& 0.2205 
 \\
|V_{23}| &=& 0.041 \pm 0.003 \\
|V_{13}/V_{23}| &=& 0.08 \pm 0.02
\end{array}
\right \}~\cite{PDG}, 
~~~
m_t^{}= 175.6 \pm 5.5~{\rm GeV}~\cite{mt96}, 
\eea
where the error of $|V_{12}|$ can be safely neglected. 
We find 
\bea
\begin{array}{l}
	\left.
	\begin{array}{lcl}
	\cosd &=& 0.36 \pm 0.83\\
	\vsk{0.2}
	R_1  &=& 0.93 \pm 0.75
	\end{array}
	\right \}~~~~ 
	\rho_{\rm corr} = 0.90.\\
\vsk{0.3}
\end{array}
\label{eq:r1_cosd}
\eea
Because of the strong positive correlation between 
the errors, only the following combination is effectively 
constrained; 
\beq
R_1 = 0.61 + 0.89 \cosd \pm 0.33.
\eeq
We show the 1-$\sigma$ (39\%) allowed region 
of $\cosd$ and $R_1$ in Fig.~\ref{bbkk39cl}. 
In the figure, there is small region which corresponds to 
$1 \le \cosd$ where the flavor mixing does not obey the CKM 
mechanism. 
\figbbkkonesigma  
The range of $\cosd$ along the $R_1=1$ line is the allowed 
region of $\cosd$ in the SM: $0.08\, \lsim \cosd\, \lsim\, 0.78$. 
We can read off from Fig.~\ref{bbkk39cl} that the current experimental 
data of $\xd$ and $\ek$ parameters constrain the new physics 
contributions within $0.18\, \lsim\, R_1\, \lsim\, 1.68$. 

Next we examine the constraint on $R_2$. 
Although the recent observation of one candidate event is unsuitable 
to include in the actual fit, we can expect that the data will be 
improved in the near future. 
In the following, we adopt the central value of the SM prediction 
as the mean value of Br($\kplusdecay$) and study consequences of 
improved measurements. 
With several more events, the branching fraction 
can be measured as 
${\rm Br}(\kplusdecay) = (0.9 \pm 0.4) \times 10^{-10}$.
Then the combined result with $\xd$ and $\ek$ parameters can be 
found as 
\bea
\begin{array}{l} \left.
	\begin{array}{lcl}
	\cosd &=& 0.36 \pm 0.83\\
	\vsk{0.2}
	R_1  &=& 0.93 \pm 0.75 \\
	\vsk{0.2}
	R_2  &=& 1.14 \pm 0.53
	\end{array}
	\right \}~~~~ 
	\rho_{\rm corr} = \left ( 
	\begin{array}{rrr}
	1 &  0.90 & 0.68  \\
	  &     1 & 0.62   \\
	  &        &   1  
		\end{array} \right ).\\
\vsk{0.3}
\end{array}
\label{eq:r1_r2}
\eea
In Fig.~\ref{sum_r1r2}, 
the results are shown on the $R_1$-$R_2$ plane for 
three values of $\cosd$; 
$\cosd = 0.36$ (mean value), $-0.47$ (${\rm mean\, value}-1\sigma$)
and 1.19 (${\rm mean\, value}+ 1\sigma$). 
\figsumr1r2 
Using this result, we can discuss about constraints on the new 
physics contributions to these processes on the $R_1$-$R_2$ plane 
for a given value of $\cosd$. 
\section{Predictions on $R_1,R_2$ in the MSSM and the THDM}
\clean
Here, we find predictions on $R_1,R_2$ in the MSSM and the THDM. 
The previous studies on those processes in both models 
can be found in \cite{BB_SUGRA, bb_mixing,
BB_goto,BB_THDM} for $\bbbar,\kkbar$ mixings, and 
\cite{susy_buras,susy_process,thdm_process} 
for $\kplusdecay$ process.  

In the MSSM based on $N=1$ supergravity~\cite{SUGRA}, 
there are several extra particles. 
Then, interactions among them could be new sources of 
FCNC processes.  
It is known that chargino --$t$-squark exchange 
and charged Higgs--$t$-quark exchange processes 
give the leading contributions to FCNC processes 
for $B$ and $K$ meson systems. 
For the chargino contribution, effects from squarks 
in the first two generations are canceled each other 
because degeneracy among their masses holds in good 
approximation. 
The interactions among the charged Higgs boson and 
the up-type quarks are the same with those of the type 
II-THDM~\cite{higgs_hunter}.  
The charged Higgs boson interacts with the up-type 
quarks through the Yukawa interactions which are 
proportional to the corresponding quark masses. 
As a result, the charged Higgs contributions to the FCNC 
processes are dominated by its interaction with the top-quark. 

The magnitudes of both the chargino and the charged Higgs 
contributions are proportional to $1/\tan\beta$, where 
$\tan\beta$ is the ratio of the vacuum expectation values 
of two Higgs fields. 
The effective Lagrangians for both contributions are 
described by $V-A$ and $S+P$ operators. 
The latter can be negligible for small $\tan\beta$. 
Furthermore, contributions from other sources in the 
MSSM do not give sizable effects to the FCNC processes 
for $\tan\beta\, \lsim\, 10$~\cite{susy_buras,BB_goto}.  
Hence we examine both models in the region $\tan\beta\, \lsim\, 10$.  

The expressions for $R_1$ in the MSSM and the THDM 
can be found in \cite{bb_mixing}. 
The MSSM contribution to the decay process 
$\kplusdecay$ is expressed by using $D^{\rm new}$ 
as follows 
\beq
D^{\rm new}(i) = \sum_{m,n,k,\alpha,\beta}
	D_C^{}(i,m,n;\ell,k;\alpha,\beta) 
	+ 	D_H^{}(i,\ell),  
\eeq
where $D_C^{}(i,m,n;\ell,k;\alpha,\beta)$  and $D_H^{}(i,\ell)$
represent the chargino and the charged Higgs boson contributions, 
respectively. Their explicit forms are given in \cite{cho_kdecay}. 
For $D_C$, by using the unitarity of the CKM matrix and the 
degeneracy of the squark masses between the first two 
generations, we obtain 
\beq
V_{i2}^* V_{i1}^{}
D_C^{}(i,m,n;\ell,k;\alpha,\beta) 
= V_{32}^* V_{31}^{} \biggl\{
D_C^{}(3,m,n;\ell,k;\alpha,\beta) 
- D_C^{}(1,m,n;\ell,k;\alpha,\beta)\biggr\}, 
\eeq
and the chargino contribution $A^{\rm new} \equiv A_C$ 
is given by 
\beq
A_C\equiv \sum_{m,n,k,\alpha,\beta}\biggl\{ 
D_C(3,m,n;\ell,k;\alpha,\beta) 
	- D_C(1,m,n;\ell,k;\alpha,\beta)\biggr\}. 
\label{eq:a_c}
\eeq

\figmssm
For $D_H$, due to the smallness of the Yukawa couplings for 
light quarks, we can write the charged Higgs contribution as 
\beq
A_H \equiv D_H(3,\ell). 
\label{eq:a_h}
\eeq
From (\ref{eq:a_c}) and (\ref{eq:a_h}), $R_2$ in the MSSM 
is defined as 
\beq
R_2 \equiv \frac{D_W^{}(3) + A_C + A_H}{D_W^{}(3)}. 
\label{mssm_r2}
\eeq
On the other hand, the THDM contribution to $R_2$ is given 
by setting $D_C = 0$ in (\ref{mssm_r2}): 
\beq
R_2 \equiv \frac{D_W^{}(3) + A_H}{D_W^{}(3)}. 
\eeq

Let us proceed numerical study. 
In order to reduce the number of input parameters in the 
MSSM, we express the soft SUSY breaking scalar masses 
in the sfermion sector by a common mass parameter 
$m_0^{}$. 
Also taking the scalar trilinear coupling $A_f^{}$ 
for sfermion $\tilde{f}$ 
as $A_f^{} = m_0^{}$, the MSSM contributions can be evaluated by 
using four parameters, $m_0^{}, \tan\beta$, the higgsino mass term 
$\mu$ and the SU(2) gaugino mass term $m_2$. 
In our study, these parameters are taken to be real. 
In Fig.~\ref{r2_mssm}, we show the 
MSSM and THDM contributions to $R_1,R_2$ parameters with 
the constraints on these parameters for $\cosd = 0.36$. 
The numerical study was performed in the range of 
$100~\gev < m_0 < 1~{\rm TeV}, |\mu| < 200~\gev$ and 
$m_2 = 200~\gev$ for $\tan\beta = 2$. 
We fixed the charged Higgs boson mass at $m_H = 200~{\rm GeV}$ 
in the MSSM prediction. 
This is the reason why the MSSM contributions do not converge 
to $R_1 = 1$ in the figure. 
We take into account the 
recent estimation of lower mass limits for lighter $t$-squark 
and lighter chargino~\cite{SUSY_mass_limit}: 
$80~{\rm GeV} \le m_{\tilde{t}1}$ and 
$91~{\rm GeV} \le m_{\chg 1}$. 
The MSSM contribution to $R_1$ interferes with that of the SM 
constructively~\cite{bb_mixing,BB_goto,kurimoto}. 
On the other hand, the contribution to $R_2$ interferes with 
that of the SM both constructively and destructively. 
Contrary to the case of the MSSM, the THDM contribution 
constructively interferes with the SM contribution for 
both $R_1$ and $R_2$. 
The Yukawa interaction between the top-quark and the 
charged Higgs boson is proportional to 
$1/\tan^2\beta$. 
Thus constraints on the THDM contribution to 
these quantities are weakened together with the increase of $\tan\beta$. 

\section{Summary}
We have studied impacts on searching for signatures of 
new physics beyond the SM from some FCNC processes -- 
$\bbbar, \kkbar$ mixings and the rare decay $\kplusdecay$.  
For a certain class of models of new physics, 
two parameters $R_1$ and $R_2$ were introduced to estimate 
the new physics contributions to $\bbbar, \kkbar$ mixings 
and $\kplusdecay$ decay, respectively. 
Then constraints on the new physics contributions are 
obtained from experimental data by using these parameters 
and $\cosd$. 

Taking account of both experimental and theoretical 
uncertainties for the $\bbbar$ and $\kkbar$ mixings, 
we found current constraint on $R_1$ as 
$0.18\, \lsim\, R_1\, \lsim\, 1.68$. 
With the assumption that the future data of Br($\kplusdecay$) 
will be close to the SM prediction, 
constraints on $\cosd,R_1$ and $R_2$ were found.  
The results were applied to the MSSM and the THDM 
contributions to those processes. 
Although there are parameter space which give roughly 50\% 
enhancement of $R_1$, contributions to $R_2$ are less 
than $\pm 10\%$. 
So quite precise experimental measurement of $\kplusdecay$ 
is required to study constraints on the parameter space
of these models. 

\section*{Acknowledgment}
The author would like to thank T. Morozumi and the organizing 
staffs of the workshop for making an opportunity to give 
his talk. 
This work is supported in part by Grant-in-Aid for Scientific 
Research from the Ministry of Education, Science and Culture 
of Japan.


\end{document}